\documentclass[11pt]{article}
\usepackage{amsfonts,amsbsy,bm,euscript,mathrsfs}
\usepackage{amssymb,stmaryrd,faktor}
\usepackage{amsmath}
\usepackage{bbm}
\usepackage{graphicx}
\usepackage[title,titletoc]{appendix}
\usepackage[bookmarks=true,colorlinks=true,linkcolor=blue,citecolor=blue,urlcolor=blue,bookmarksnumbered]{hyperref}
\usepackage{dsfont}
\usepackage{lmodern}
\usepackage{mathrsfs}
\usepackage{mathtools}
\usepackage{bbm}
\usepackage{braket}
\usepackage{slashed}
\usepackage[normalem]{ulem}
\usepackage{booktabs}
\usepackage{subfig}
\usepackage{tikz}
\usepackage[utf8]{inputenc}
\usepackage[english]{babel}
\usepackage{a4wide}
\usepackage{cite}
\usepackage{appendix}
\usepackage{etoolbox}
\usepackage{physics}
\usepackage{sectsty}
\usepackage{float}
\usepackage{systeme}
\usepackage{xcolor}
\usepackage{mathdots}

\DeclareMathOperator{\dif}{d\!}
\DeclareMathOperator{\e}{e}
\DeclareMathOperator{\im}{i\!}
\DeclareMathOperator{\alg}{\mathfrak{g}}
\DeclareMathOperator{\algf}{\mathfrak{f}}
\DeclareMathOperator{\algl}{\mathfrak{l}}

\DeclareMathOperator{\gl}{\mathfrak{gl}(1|1)}
\DeclareMathOperator{\psuR}{\mathfrak{psu}(1|1)\ltimes\mathbb{R}^{3}}

\DeclareMathOperator{\psudosR}{\mathfrak{psu}(1|1)^2\ltimes\mathbb{R}^{3}}

\DeclareMathOperator{\Qg}{\mathfrak{Q}}
\DeclareMathOperator{\Pg}{\mathfrak{P}}
\DeclareMathOperator{\Sg}{\mathfrak{S}}
\DeclareMathOperator{\Kg}{\mathfrak{K}}
\DeclareMathOperator{\Hg}{\mathfrak{H}}
\DeclareMathOperator{\Cg}{\mathfrak{C}}
\DeclareMathOperator{\Dg}{\mathfrak{D}}
\DeclareMathOperator{\Xg}{\mathfrak{X}}
\DeclareMathOperator{\Jg}{\mathfrak{J}}
\DeclareMathOperator{\Cs}{\mathbb{C}^{1|1}}
\DeclareMathOperator{\Ls}{\mathit{L}^2(\mathrm{S}^1)}
\DeclareMathOperator{\inn}{\mathrm{in}}
\DeclareMathOperator{\outt}{\mathrm{out}}
\DeclareMathOperator{\opp}{\mathrm{op}}

\def\texto
       {\topmargin -20pt
        \oddsidemargin 0pt
        \headheight 0pt
        \headsep 0pt
        \textwidth 6.35in
        \textheight 9.25in
        }
\texto

\def\baselinestretch{1.2}
\sectionfont{\large}
\renewcommand{\theequation}{\thesection.\arabic{equation}}
\csname @addtoreset\endcsname{equation}{section}

\begin{document} 

\begin{titlepage}

\begin{center}
\hfill DMUS-MP-23/07

\hfill ZMP-HH/23-4

\phantom{\LARGE X}

\vskip 0.25in

{\large \textbf{Infinite-dimensional R-matrices for the relativistic scattering of massless modes on $\boldsymbol{\mathrm{AdS}_2}$}}

\vskip 0.45in

\textbf{Juan Miguel Nieto Garc\'ia,\footnote{{\texttt{juan.miguel.nieto.garcia@desy.de}}}$^{a}$ Roberto Ruiz,\footnote{{\texttt{roberto.ruiz@ift.csic.es}}}$^{b}$ Alessandro Torrielli,\footnote{{\texttt{a.torrielli@surrey.ac.uk}}}$^{c}$ 
\setcounter{footnote}{0}}

\vskip 0.15in

$\vphantom{x}^{a}$\! II. Institut für Theoretische Physik,\\
Universität Hamburg, \\
Luruper Chaussee 149, \\
22761 Hamburg, Germany \\ 

\vskip 0.15in

$\vphantom{x}^{b}$\! Instituto de F{\'i}sica Te{\'o}rica UAM/CSIC \\
C/ Nicolás Cabrera 13--15 \\
Universidad Aut{\'o}noma de Madrid \\
Cantoblanco, Madrid 28049, Spain \\ 

\vskip 0.15in
$\vphantom{x}^{c}$\! Department of Mathematics, School of Mathematics and Physics\\University of Surrey, Guildford, GU2 7XH

\vskip 0.1in 

\end{center}

\vskip .5in

\centerline{\bf Abstract}

\noindent 
We construct infinite-dimensional R-matrices that generalise the relativistic scattering of massless modes with the same chirality on $\mathrm{AdS}_2$ near the Berestein-Maldacena-Nastase vacuum. We show that the infrared limit of the R-matrices reproduces finite-dimensional scattering of massless modes on $\mathrm{AdS}_2$, from which the R-matrices borrow modified braiding unitary. We also prove that the R-matrices enjoy an infinite-dimensional symmetry superalgebra that embeds that of~$\mathrm{AdS}_2$. Finally, we verify that the R-matrices are also invariant under crossing symmetry.

\vskip .1in

\begin{sloppypar}

\noindent

\end{sloppypar}

\vskip .4in

\vskip .1in

\noindent

\end{titlepage} 

\def\baselinestretch{1}
\baselineskip 15pt
\sectionfont{\large} 
\renewcommand{\theequation}{\thesection.\arabic{equation}} 
\csname @addtoreset\endcsname{equation}{section}

%%%%%%%%%%%%%%%%%%%%%%%%%%%%%%%%%%%%%%%%%%%%%%%%%%%%%%%%%%%%%%%%%%%%%%%%%%
%%%%%%%%%%%%%%%%%%%%%%%%%%%%%%%%%%%%%%%%%%%%%%%%%%%%%%%%%%%%%%%%%%%%%%%%%%

\tableofcontents

\section{Introduction} 

The propagation of superstrings on $\mathrm{AdS}_2\times\mathrm{S}^2$-backgrounds have aroused interest since the dawn of the proposition of the AdS/CFT correspondence~\cite{9711200,9905111}. The $\mathrm{AdS}_{2}/\mathrm{CFT}_{1}$ correspondence is believed to connect superstring theory on $\mathrm{AdS}_2\times\mathrm{S}^2\times\mathrm{T}^{6}$ ---$\mathrm{AdS}_2$ for short--- to either planar superconformal quantum mechanics or chiral $\mathrm{CFT}_{2}$, and research to uncover this duality spanned decades by now \cite{ads24, dual3, dual4, gen, gen1, gen2, gen3, gen4, gen5, gen6, gen7, gen9, gen10, gen11, gen12, gen13, gen14, gen15,Aniceto:2020saj,Aniceto:2021xhb,2212.09139}. 

One early insight was the identification of the classical non-linear $\sigma$-model on $\mathrm{AdS}_2\times\mathrm{S}^2$ with the supercoset model on $\mathrm{PSU}(1,1|2)/\mathrm{SO}(1,1) \times\mathrm{SO}(2)$ \cite{Metsaev:1998it1, Metsaev:1998it2}, to which the Green-Schwarz action on $\mathrm{AdS}_2$ admits a consistent truncation \cite{Sorokin:2011rr}. Despite the truncation, target-space fermions along the non-coset directions of $\mathrm{AdS}_2$ can be neither gauged away nor decoupled in general. This fact implies that classical superstrings on $\mathrm{AdS}_2$ do not reduce to a decoupled supercoset and flat directions \cite{Sorokin:2011rr}. Classical superstrings are integrable nonetheless, at least to second order in fermions \cite{Sorokin:2011rr}, and their classical integrability still relies upon a $\mathbb{Z}_{4}$-automorphism \cite{Cagnazzo:2011at}.

The understanding of classical superstrings on $\mathrm{AdS}_2$ laid the foundations for the computation of their quantum spectrum. For instance, finite-gap equations, based on classical integrability, provided an educated guess for the asymptotic Bethe equations \cite{Sorokin:2011rr}. One further advance was the postulation of the exact quantum S-matrix of $\mathrm{AdS}_2$~\cite{Hoare:2014kma}. The proposal mimicked the non-relativistic world-sheet S-matrix of $\mathrm{AdS}_{5}\times\mathrm{S}^{5}$ \cite{BeisReview}. The S-matrix is based on the centrally extended superalgebra $\psudosR$. This superalgebra is the proposal for the all-loop completion of the perturbative symmetry superalgebra that the S-matrix above Berenstein-Maldacena-Nastase (BMN) vacuum preserves \cite{amsw, amsw1, amsw2, Per6, Per10}. The superalgebra $\psudosR$, for what concerns its central extension, is not derived but postulated, and, thus, there is enough room, at least in principle, for other superalgebras to realise non-perturbative symmetries of the S-matrix beyond all-loop.

Despite being non-relativistic, the exact S-matrix of $\mathrm{AdS}_2$ scatters both massive and massless modes~\cite{Hoare:2014kma}. Massive modes on $\mathrm{AdS}_2$ are unconventional, in the sense that they belong to long representations of $\psudosR$. No shortening condition fixes the masses, which are therefore just parameters of massive representations. Massive modes enjoy Yangian symmetry \cite{Hoare:2014kma}, but the evaluation representation must be non-standard \cite{Hoare:2014kmaa}. Moreover, the S-matrix of massive modes is crossing-symmetric and unitary.  Although the S-matrix is fairly complicated and the dressing factor remains to be fully determined, the analysis of the massive sector seems to be on the right track because there is an agreement with perturbative computations at the first few orders holds \cite{amsw, amsw1, amsw2, Per6, Per10}.

As opposed to massive modes, massless modes on $\mathrm{AdS}_2$ belong to short representations of~$\psudosR$. The dispersion relation of massless modes follows from a shortening condition. Massless representations enjoy Yangian symmetry, whose realisation is canonical, as opposed to the situation of massive representations \cite{Hoare:2014kma,Hoare:2014kmaa}.  Although the S-matrix of massless modes is the massless limit of the S-matrix for massive modes \cite{Hoare:2014kma}, it exhibits the traditional mismatching with perturbative computations \cite{amsw1, amsw2, Per6, Per10}. This mismatching shows that the S-matrix of massless modes is non-perturbative, which indicates that the situation of $\mathrm{AdS}_{2}$ is analogous to that of $\mathrm{AdS}_{3}$  \cite{Borsato:2016xns,DiegoBogdanAle,Cavaglia:2021eqr,Ekhammar:2021pys,Cavaglia:2022xld,Brollo:2023pkl}. The non-perturbative nature is prominently featured near the BMN vacuum, where a relativistic limit of the S-matrix controls scattering \cite{Andrea2}. The expectation based on perturbative computations is that the scattering of massless modes, which propagate at the world-sheet speed of light, do not scatter unless they carry opposite chirality. However, at the point where transition amplitudes are otherwise trivial, the relativistic limit of the S-matrix between massless modes of the same chirality is precisely non-diagonal \cite{Andrea2}. This fact ---also encountered in $\mathrm{AdS}_{3}$ \cite{DiegoBogdanAle}--- relates to massless modes in relativistic integrable $\mathrm{QFT}_2$ \cite{Zamol2,Zamol21}. Moreover, this S-matrix also enjoys ${\cal{N}}=1$ supersymmetry, and, in fact, arises in the massless and ultrarelativistic limit of standard S-matrices in integrable $\mathcal{N}=1$ $\mathrm{QFT}_2$ \cite{Fendley:1990cy,Schoutens,MC}. There exists a proposal for the dressing factor of $S$ \cite{Andrea2}, whose properties have been studied in detail \cite{AA2,DVA}.

The R-matrix associated to the relativistic limit of the S-matrix between massless modes of the same chirality satisfies the Yang-Baxter (YB) equation. It is an eight-vertex R-matrix, the signature of quantum integrability of the eight-vertex model, from which the R-matrix borrows name, and the XYZ spin chain \cite{Baxter:1972hz, Baxter:1972hz1}. Being an eight-vertex R-matrix, it lacks a standard pseudo-vacuum, thus rendering the algebraic Bethe ansatz inapplicable. Nonetheless, the S-matrix fulfils the free-fermion condition, which, in line with $\mathcal{N}=1$ $\mathrm{QFT}_2$ \cite{Ahn:1993qa}, facilitates a solution to this problem \cite{Andrea2}. The free-fermion condition enables, in particular, the application of inversion relations \cite{Zamolodchikov:1991vh}. Provided the eigenvalues of the appropriate transfer matrix are known, inversion relations yield a set of Bethe equations that are comparable with the equations based on superstring theory \cite{Sorokin:2011rr}. This state of affairs motivated the brute-force diagonalisation of the transfer matrix with a few massless modes, and the clarification of Yangian symmetry \cite{alea}. Efforts culminated in the general analysis of the free-fermion condition in the AdS/CFT correspondence~\cite{Marius}, which turns out to be a ubiquitous trait of superstring R-matrices. The analysis revealed that a Bogoliubov transformations permits the diagonalisation of the transfer matrix in the fashion of the transfer matrix of free fermions. The Bogoliubov transformation in $\mathrm{AdS}_2$ enabled the construction of a state akin to the pseudo-vacuum (the `pseudo-pseudo-vacuum') for the transfer matrix of two massless modes \cite{Marius}.  

These results came along with parallel developments that are worth mentioning. First, the change of variables that recasts the S-matrices of massless modes in the so-called difference form \cite{AA2}. The change of variables ---also used in $\mathrm{AdS}_{3}$ \cite{gamma2,AleSSergey,AleSSergey2,AleSSergey3}--- actually admits a generalisation to massive modes, which partially puts the S-matrix in difference form \cite{AleSSergey,AleSSergey2,AleSSergey3}. Second, the construction of an S-matrix that interpolates between $\mathrm{AdS}_2$ and $\mathrm{AdS}_3$ \cite{Marius2},~\footnote{We thank Ben Hoare for very useful discussions.}  based itself on previous work \cite{Dublin,Dublin2,Dublin3,Dublin4,Dublin5}, and the determination of the associated symmetry superalgebra.

In this article, we will construct infinite-dimensional R-matrices that generalise the relativistic scattering of massless modes with the same chirality on $\mathrm{AdS}_2$ near the BMN vacuum. In section~\ref{salg}, we will present the embedding $\alg$ of the superalgebra $\psuR$ of $\mathrm{AdS}_2$, which will play the role of zeroth level of symmetry superalgebra of the R-matrices. We will also put forward the infinite-dimensional representation of $\alg$ with which we will be concerned. In section \ref{sol}, on the basis of both heuristics around the Khoroshkin-Tolstoy (KT) formula of $\mathfrak{gl}(1|1)$ and $\alg$, we will solve the YB equation and obtain the infinite-dimensional R-matrices we will focus on. In section~\ref{sads2}, we will define the infrared (IR) limit of the R-matrices that reproduces the finite-dimensional R-matrices of \cite{Andrea2}. In section \ref{bus}, we will fix the free coefficients that infinite-dimensional R-matrices comprise in order to satisfy modified braiding unitarity as defined in \cite{Andrea2}. In section \ref{ssym}, we will compute the symmetry superalgebra of the R-matrices. We will prove that the invariance of the R-matrices under $\alg$ will force us to embed $\alg$ into an infinite-dimensional superalgebra $\algl$. We will also verify that the R-matrices are relativistic invariant.  In section~\ref{sRcross}, we will use the antipode of $\algl$ to construct the antiparticle representation of $\algl$ and prove that the R-matrices fulfil crossing symmetry. In section \ref{sconclusions}, we will conclude with some general remarks and prospects on future research.

\section{The embedding of \texorpdfstring{$\boldsymbol{\mathfrak{psu}(1|1)\ltimes\mathbb{R}^3}$}{psu(11)R3} and its infinite-dimensional representation}

\label{salg}

Reference \cite{Hoare:2014kma} proposed $\psudosR$ to be the all-loop completion of the symmetry superalgebra preserved by the S-matrix above the BMN vacuum. The symmetry superalgebra of massless modes with the same chirality, left or right, is in particular $\psuR$. Reference \cite{Andrea2} showed that $\psuR$ is realised in the relativistic limit of massless representations at vanishing superstring tension. In this section, we will introduce the embedding $\alg$ of $\psuR$, which will play the role of the infinite-dimensional zeroth-level of symmetry superalgebra of the R-matrix. We will specify $\alg$ in infinite-dimensional representations that will generalise the representations of $\psuR$ that \cite{Andrea2} put forward.~\footnote{We will avoid notation that discriminates between the superalgebra and representation for brevity. The meaning of the notation will be clear from the context.}

Our starting point is $\mathfrak{psl}(1|1){\ltimes}\mathbb{C}^{3}$, the maximal central extension of  $\mathfrak{psl}(1|1)$. The pair of fermions $\Qg$ and $\Sg$, and the central extensions $\Pg$, $\Kg$, and $\Hg$ span the superalgebra. If $\Pg$, $\Kg$, and $\Hg$ were absent, $\Qg$ and $\Sg$ would span $\mathfrak{psl}(1|1)$; the introduction of $\Hg$ alone would promote $\mathfrak{psl}(1|1)$ to $\mathfrak{sl}(1|1)$. The non-trivial (anti-)commutation relations of $\mathfrak{psl}(1|1){\ltimes}\mathbb{C}^{3}$ read
\begin{equation}
\label{compsu}
\{\Qg,\Qg\}=2\Pg \ , \quad \{\Sg,\Sg\}=2\Kg \ , \quad \{\Qg,\Sg\}=2\Hg \ .
\end{equation}
The superalgebra is compatible with the Lorentz boost $\Jg$, whose non-trivial trivial commutation relations reads
\begin{equation}
\label{Jcomm}
[\Jg , \Qg]=\frac{1}{2} \Qg \ , \quad [\Jg , \Sg]=\frac{1}{2} \Sg \ , \quad [\Jg , \Pg]= \Pg \ , \quad [\Jg , \Kg]=\Kg \quad  \quad [\Jg , \Hg]= \Hg \ .
\end{equation}
The action of $\Jg$ through (\ref{Jcomm}) defines an outer automorphism of $\mathfrak{psl}(1|1){\ltimes}\mathbb{C}^{3}$ onto itself, under which $\Qg$ and $\Sg$ carry the weight one-half, and $\Pg$, $\Kg$, and $\Hg$ carry weight one.~\footnote{We should emphasise that $\Jg$ is not the boost $\mathcal{J}$ of \cite{AA2}, but the outer automorphism $\mathcal{D}$. The action of $\mathcal{J}$ on $\psuR$ is ill-defined in the relativistic limit of one-particle massless representations.} Moreover, we will reduce $\mathfrak{psl}(1|1){\ltimes}\mathbb{C}^{3}$ to $\mathfrak{psu}(1|1){\ltimes}\mathbb{R}^{3}$ 
by the imposition of
\begin{equation}
\label{unitarity}
\Qg^{\dagger}=\im\Sg \ , \quad \Pg^{\dagger}=-\Kg \ , \quad \Hg^{\dagger}=\Hg \ ,
\end{equation}
at the level of  representations.

The one-particle massless representation of $\mathfrak{psl}(1|1){\ltimes}\mathbb{C}^{3}$ in the relativistic limit of \cite{Andrea2} acts on $\Cs$.~\footnote{
\label{fn}
Reference \cite{Andrea2} actually put forward two one-parameter families of such representations, namely the right and left families. Each family is parameterised by $\alpha$. We will focus on the right representation with $\alpha=1$ for definiteness. Our analysis extends to the remaining cases without major modifications.} Let
\begin{equation}
\left\{\ket{\varphi}=\begin{bmatrix}
1 \\
0 \\
\end{bmatrix} \ , \quad
\ket{\psi}=\begin{bmatrix}
0 \\
1 \\
\end{bmatrix}
\right\} \ ,
\end{equation}
be the linear basis of $\Cs$ that the boson $\ket{\varphi}$ and the fermion $\ket{\psi}$ span.  The representation reads
\begin{alignat}{4}
\label{repmat}
    \Qg=\frac{1}{\sqrt{2}}\e^{\theta/2-\im\pi/4} 
    \begin{bmatrix} 
    0 & 1 \\ 
    1 & 0  
    \end{bmatrix} &\ , \quad&
    \Pg=-\frac{\im}{2}\e^{\theta} 
    \begin{bmatrix} 
    1 & 0 \\
    0 & 1  
    \end{bmatrix} &\ , \quad & 
    \Hg=\frac{1}{2}\e^{\theta} 
    \begin{bmatrix} 
    1 & 0 \\
    0 & 1  
    \end{bmatrix} \ , \nonumber \\
    \Sg=\frac{1}{\sqrt{2}}\e^{\theta/2+\im\pi/4} 
    \begin{bmatrix} 
    0 & 1 \\
    1 & 0  
    \end{bmatrix} &\ , \quad &
    \Kg=\frac{\im}{2}\e^{\theta} 
    \begin{bmatrix} 
    1 & 0 \\
    0 & 1  
    \end{bmatrix} & \ . 
\end{alignat}
where $\theta$ denotes the relativistic rapidity of one particle (related to the relativistic world-sheet momentum like $p=\exp(\theta)$). Condition (\ref{unitarity}) holds at $\theta\in\mathbb{R}+\im\pi\mathbb{Z}$. The representation (\ref{repmat}) is short since the shortening condition
\begin{equation}
\Hg^2-\Pg\Kg=0 \ ,
\end{equation}
holds. Finally, the representation of $\Jg$ is
\begin{equation}
\label{Jrep}
\Jg=\frac{\partial}{\partial\theta}
    \begin{bmatrix} 
    1 & 0 \\
    0 & 1  
    \end{bmatrix} \ .
\end{equation}

To define the superalgebra $\alg$, we introduce the boson $\Cg$ first, which will in turn lead us to the representation we are interested in. In absence of $\Pg$ and $\Kg$,  $\Cg$ is the Cartan generator of $\gl$,  the superalgebra that $\Cg$, $\Qg$, $\Sg$, and $\Hg$ span. The superalgebra $\gl$ has non-trivial (anti-)commutation relations 
\begin{equation}
\label{comgl}
    [\Cg,\Qg]=-2\Qg \ , \quad [\Cg,\Sg]=2\Sg \ , \quad \{\Qg,\Sg\}=2\Hg \ .
\end{equation}
The boson $\Cg$ has zero weight under the action of $\Jg$:
\begin{equation}
\label{JC}
[\Jg,\Cg]=0 \ .    
\end{equation}
In the presence of $\Pg$ and $\Kg$, consistency with the super-Jacobi-identity in $\mathfrak{g}$ implies that 
\begin{equation}
\label{comCPK}
     [\Cg,\Pg]=-4\Pg \ , \quad [\Cg,\Kg]=4\Kg \ , 
\end{equation}
must hold in addition to (\ref{compsu}) and (\ref{comgl}). In $\mathfrak{g}$, $\Kg$ and $\Pg$ are not central any more. We deduce that we cannot write $\Cg$ in the representation (\ref{repmat}) because $\Pg$ and $\Kg$ in (\ref{repmat}) are proportional to the identity on $\Cs$.

To make the representation compatible with $\Cg$, we promote (\ref{repmat}) to a representation over the super-Hilbert space $\mathcal{H}=\Cs\otimes L^2(\mathrm{S}^{1})$, where $L^2(\mathrm{S}^{1})$ lacks any fermionic grading. The new representation is infinite-dimensional. In position space, this representation reads 
\begin{alignat}{4}
\label{reppos}
    \Qg=\frac{1}{\sqrt{2}}\e^{-2\im\phi/L+\theta/2-\im\pi/4} 
    \begin{bmatrix} 
    0 & 1 \\ 
    1 & 0  
    \end{bmatrix} &\ , \quad&
    \Pg=-\frac{\im}{2}\e^{-4\im\phi/L+\theta} 
    \begin{bmatrix} 
    1 & 0 \\
    0 & 1  
    \end{bmatrix}
     &\ ,  \quad&
    \Hg=\frac{1}{2}\e^{\theta} 
    \begin{bmatrix} 
    1 & 0 \\
    0 & 1  
    \end{bmatrix} &\ ,& \nonumber \\
    \Sg=\frac{1}{\sqrt{2}}\e^{2\im\phi/L+\theta/2+\im\pi/4} 
    \begin{bmatrix} 
    0 & 1 \\
    1 & 0  
    \end{bmatrix} &\ , \quad&
    \Kg=\frac{\im}{2}\e^{4\im\phi/L+\theta} 
    \begin{bmatrix} 
    1 & 0 \\
    0 & 1  
    \end{bmatrix} &\ , \quad&
    \Cg=-\im L\frac{\partial}{\partial\phi}
    \begin{bmatrix} 
    1 & 0 \\
    0 & 1  
    \end{bmatrix} &\ .&
\end{alignat}
where the angle $\phi\in[-\pi L,\pi L)$ parameterises a circle manifold $\mathrm{S}^{1}$ of radius $L$. The representation of $\Jg$ on $\mathcal{H}$ is still (\ref{Jrep}). Position space makes clear that we have identified~$\Cg$ with the momentum operator over $\Ls$. Conditions (\ref{unitarity}) and $\Cg=\Cg^{\dagger}$ hold for this representation at $\theta\in\mathbb{R}+\im\pi\mathbb{Z}$, just as before.

Although we have motivated (\ref{reppos}) in position space, the momentum space is better suited to write the R-matrix. Let us introduce the orthonormal basis of plane waves of~$\Ls$:
\begin{equation}
\label{pw}
\left\{\ket{n}:\bra{\phi}\ket{n}=\frac{1}{\sqrt{2\pi L}}\e^{\im n \phi/L} \ , \ n\in\mathbb{Z}\right\} \ .
\end{equation}
The integer momentum $n$ labels plane waves, and we call $\Ls$ `momentum space' for this reason.
The representation (\ref{reppos}) in momentum space reads
\begin{alignat}{4}
\label{repmom}
    \Qg=\frac{1}{\sqrt{2}}\e^{\theta/2-\im\pi/4} 
    \begin{bmatrix} 
    0 & b \\ 
    b & 0  
    \end{bmatrix} &\ , \quad&
    \Pg=-\frac{\im}{2}\e^{\theta} 
    \begin{bmatrix} 
    b^2 & 0 \\
    0 & b^2  
    \end{bmatrix}
     &\ ,  \quad&
    \Hg=\frac{1}{2}\e^{\theta} 
    \begin{bmatrix} 
    1 & 0 \\
    0 & 1  
    \end{bmatrix} &\ ,& \nonumber \\
    \Sg=\frac{1}{\sqrt{2}}\e^{\theta/2+\im\pi/4} 
    \begin{bmatrix} 
    0 & b^{\dagger} \\
    b^{\dagger} & 0  
    \end{bmatrix} &\ , \quad&
    \Kg=\frac{\im}{2}\e^{\theta} 
    \begin{bmatrix} 
    (b^{\dagger})^2 & 0 \\
    0 & (b^{\dagger})^2 
    \end{bmatrix} &\ , \quad&
    \Cg=
    \begin{bmatrix} 
    N & 0 \\
    0 & N  
    \end{bmatrix} &\ ,&
\end{alignat}
where $b$, $b^{\dagger}$, and $N$ are operators on $\Ls$ whose action reads
\begin{equation}
b\ket{n}=\ket{n-2} \ , \quad b^{\dagger}\ket{n}=\ket{n+2} \ , \quad N\ket{n}=n\ket{n} \ .
\end{equation}
Despite the superficial analogy with the quantum harmonic oscillator, we must underline that the kernel of $b$ is empty and $b^{-1}=b^{\dagger}$. 

We are now in a position to define $\alg$, the superalgebra that $\psuR$ together with the bosons $\Cg$, $\Dg$ and $\Dg^{-1}$ build up. The introduction of $\Dg$ and $\Dg^{-1}$ cannot be justified at this stage. Suffice to say, we will need that $\Dg$ and $\Dg^{-1}$ belong to $\alg$ to construct the zeroth-level of the infinite-dimensional symmetry superalgebra of the R-matrix. The only non-trivial commutation relation of $\Dg$ and~$\Dg^{-1}$ inside $\alg$ is
\begin{equation}
\label{CD}
[\Cg,\Dg]=-2\Dg \ ,  \quad [\Cg,\Dg^{-1}]=2\Dg^{-1}
\end{equation}
whereas they have zero weight under the action of $\Jg$:
\begin{equation}
\label{JD}
[\Jg,\Dg]=0 \ , \quad [\Jg,\Dg^{-1}]=0 \ .
\end{equation}
The representation of $\Dg$ on $\mathcal{H}$ is 
\begin{equation}
\label{repD}
\Dg=
    \begin{bmatrix} 
    b & 0 \\ 
    0 & b
    \end{bmatrix} 
\ , \quad \Dg^{-1}=
    \begin{bmatrix} 
    b^{\dagger} & 0 \\ 
    0 & b^{\dagger}
    \end{bmatrix} 
\ .
\end{equation}

\section{The Yang-Baxter equation}

\label{sol}

In this section, we will construct the R-matrices on which we will focus on hereafter. By definition, an R-matrix is a linear operator $R:\mathcal{H}\otimes\mathcal{H}\rightarrow\mathcal{H}\otimes\mathcal{H}$ that satisfies the YB equation
\begin{equation}
\label{YB}
R_{12}R_{13}R_{23}=R_{23}R_{13}R_{12} \ ,
\end{equation}
where~\footnote{We will use 1 to denote either one, the identity operator, and the identity on the infinite-dimensional algebra we introduce in section \ref{ssym}. The meaning of 1 will be clear from the context.}
\begin{equation}
R_{12}=R\otimes 1 \ , \quad  
R_{23}=1 \otimes R \ , \quad 
R_{13}=(\Pi \otimes 1)R_{23}(\Pi \otimes 1) \ ;
\end{equation}
$1:\mathcal{H}\rightarrow\mathcal{H}$ denotes the identity operator; and $\Pi:\mathcal{H}\otimes\mathcal{H}\rightarrow\mathcal{H}\otimes\mathcal{H}$ denotes the graded permutation, defined by
\begin{equation}
    \label{gradp}
    \begin{split}
    \Pi\ket{\varphi,m,\varphi,n}=\ket{\varphi,n,\varphi,m} \ , \quad  \Pi\ket{\varphi,m,\psi,n}&=\ket{\psi,n,\varphi,m} \ , \\ 
\Pi\ket{\psi,m,\varphi,n}=\ket{\varphi,n,\psi,m} \ , \quad \Pi\ket{\psi,m,\psi,n}&=-\ket{\psi,n,\psi,m} \ .
\end{split}
\end{equation}
We approach the resolution of (\ref{YB}) by means of an ansatz for $R$. To write the ansatz, we rearrange $\mathcal{H}\otimes\mathcal{H}\cong(\Cs\otimes\Cs)\otimes(\Ls\otimes\Ls)$ first. We then rewrite $R$ as a $(4\times4)$ supermatrix whose entries are linear operators on the momentum space of two particles. Based on heuristics around the KT formula for $\gl$ we comment on below, the ansatz we propose is
\begin{equation}
\label{ansatz}
R_{\pm}=d_{\pm}\,\Sigma_{\alpha}\otimes\Sigma_{\beta}+a_{\pm}\Sigma_{\gamma}\otimes\Sigma_{\delta}\ ,
\end{equation}
where $\alpha,\beta=0,3$; $\gamma,\delta=1,2$;
\begin{equation}
\Sigma_{0}=
\begin{bmatrix}
1 & 0 \\
0 & 1
\end{bmatrix} \ , \quad 
\Sigma_{1}=
\begin{bmatrix}
0 & 1 \\
1 & 0
\end{bmatrix} \ , \quad 
\Sigma_{2}=
\begin{bmatrix}
0 & -1 \\
1 & 0
\end{bmatrix} \ , \quad 
\Sigma_{3}=
\begin{bmatrix}
1 & 0 \\
0 & -1
\end{bmatrix} \quad  \ ;
\end{equation} 
and
\begin{equation}
\label{daans}
\begin{split}    
d_{\pm}\ket{m,n}&=\overset{\infty}{\underset{N=0}{\sum}}\delta_N\ket{m\mp4N,n\pm4N} \ , \\
a_{\pm}\ket{m,n}&=\overset{\infty}{\underset{N=0}{\sum}}\alpha_N \, \ket{m\mp4N\mp2,n\pm4N\pm2} \ ,
\end{split}
\end{equation}
where $\delta_{N}=\delta_{N}(\theta_1,\theta_2)$, $\alpha_{N}=\alpha_{N}(\theta_1,\theta_2)$, and $\theta_1$ and $\theta_2$ are the relativistic rapidities of the first and  second particles, respectively. We can replace the graded tensor product $\otimes$ by the regular tensor product in (\ref{ansatz}) because both determine the same set of $(4\times4)$ supermatrices. 

As we have already mentioned, the ansatz (\ref{ansatz}) is reminiscent of the KT formula of $\gl$, the closed formula for the universal R-matrix of the Yangian double of $\gl$ \cite{KTform}. Specifically, we base~(\ref{ansatz}) on~$\mathcal{R}_\pm$, the factors that comprise the fermions in the KT formula \cite{KTform}. Let us comment on the heuristics behind this approach. In the evaluation representation, 
\begin{equation}
\label{KT}
\mathcal{R}_+\sim\underset{N=0}{\overset{\infty}{\sum}} \beta_N \Qg^N \otimes \Sg^N \ , \quad \mathcal{R}_-\sim\underset{N=0}{\overset{\infty}{\sum}} \beta_N \Sg^N \otimes \Qg^N \ ,
\end{equation}
where $\beta_N$ denote certain coefficients that depend on the spectral parameter of the evaluation representation, and $\Qg$ and $\Sg$ are the fermions of $\gl$. Now, $\Qg$ and $\Cg$ admit the representation~(\ref{repmom}). The substitution of (\ref{repmom}) into~(\ref{KT}) leads to expressions for $\mathcal{R}_\pm$ that falls into the class of supermatrices of (\ref{ansatz}). The pair $d$ and $a$ in~(\ref{daans}) mimic the action of the series (\ref{KT}), which split into even and odd $N$, on states of the momentum space of two particles. The reason for the dependence of $\beta_N$ on $\theta$ is that \cite{2203.15367} formally retrieved solutions 3 and 5 of \cite{Andrea2} in the ultrarelativistic limit of the KT formula of $\gl$ under the identification of the spectral parameter and the relativistic world-sheet momentum.~\footnote{We should emphasise, however, that the result of the application of the relativistic limit of \cite{Andrea2} to the spectral parameter of \cite{Hoare:2014kmaa} does not depend on $\theta$, hence the identification cannot hold without modification.} Despite motivations, we emphasise that the resemblance between~(\ref{ansatz}) and the KT formula of $\gl$ is just a faint echo. The use of~(\ref{repmom}) in the evaluation representation of the actual KT formula of~$\gl$ ---see, for instance, formulae (4.1)--(4.7) of \cite{2203.15367}--- simply fails to satisfy the YB equation (\ref{YB}). 

Let us come back to the ansatz (\ref{ansatz}). If we introduce the ansatz into the YB equation (\ref{YB}), we obtain that the only possible solutions are
\begin{equation}
\label{mats1}
R_{\pm,1}=
\begin{bmatrix}
{d_{\pm}} & 0 & 0 & {a_{\pm}} \\
0 & {d_{\pm}} & {a_{\pm}} & 0 \\
0 & {a_{\pm}} & -{d_{\pm}} & 0 \\
{a_{\pm}} & 0 & 0 & -{d_{\pm}}
\end{bmatrix} 
\ , \quad R_{\pm,2}=
\begin{bmatrix}
{d_{\pm}} & 0 & 0 & {a_{\pm}} \\
0 & -{d_{\pm}} & {a_{\pm}} & 0 \\
0 & {a_{\pm}} & {d_{\pm}} & 0 \\
{a_{\pm}} & 0 & 0 & -{d_{\pm}}
\end{bmatrix} 
\ , \\
\end{equation}
if the condition
\begin{equation}
\label{dad1}
(d_\pm)_{12}(a_\pm)_{13}(d_\pm)_{23}=(a_\pm)_{23}(d_\pm)_{13}(a_\pm)_{12} \ ,
\end{equation}
holds, and
\begin{equation}
\label{mats2}
R_{\pm,3}=
\begin{bmatrix}
{d_{\pm}} & 0 & 0 & {a_{\pm}} \\
0 & {d_{\pm}} & -{a_{\pm}} & 0 \\
0 & -{a_{\pm}} & -{d_{\pm}} & 0 \\
{a_{\pm}} & 0 & 0 & -{d_{\pm}}
\end{bmatrix} 
\ , \quad R_{\pm,4}=
\begin{bmatrix}
{d_{\pm}} & 0 & 0 & {a_{\pm}} \\
0 & -{d_{\pm}} & -{a_{\pm}} & 0 \\
0 & -{a_{\pm}} & {d_{\pm}} & 0 \\
{a_{\pm}} & 0 & 0 & -{d_{\pm}}
\end{bmatrix} 
\ , \\
\end{equation}
if the condition
\begin{equation}
\label{dad2}
(d_\pm)_{12}(a_\pm)_{13}(d_\pm)_{23}=-(a_\pm)_{23}(d_\pm)_{13}(a_\pm)_{12} \ ,   
\end{equation}
holds. To write (\ref{dad1}) and (\ref{dad2}), we have used that $d_{\pm}$ and $a_{\pm}$ commute by construction, and that $\Pi$ is just the regular permutation on the momentum space of two particles. If we flip $a_{\pm}\mapsto-a_{\pm}$ in~(\ref{mats2}) and (\ref{dad2}), we deduce that $R_{\pm,a}$ solve the YB equation if and only if
\begin{equation}
\label{eqLMN}
\delta_{L}(\theta_1,\theta_2)\alpha_{M}(\theta_1,\theta_3)\delta_{N}(\theta_2,\theta_3)=
\alpha_{N}(\theta_2,\theta_3)\delta_{M}(\theta_1,\theta_3) \alpha_{L}(\theta_1,\theta_2)\ ,
\end{equation}
for every $L, M, N >0$. Equation (\ref{eqLMN}) admits the general solution
\begin{equation}
\label{alphadelta}
\alpha_{N}=\exp(\gamma\theta)\delta_{N}  \ ,
\end{equation}
 where $\gamma$ is a constant, $\theta=\theta_1-\theta_2$ denotes the difference of relativistic rapidities, and $\delta_{N}$ are arbitrary functions. 
 
 Equation (\ref{alphadelta}) determines a solution to the YB equation, whose associated two-body S-matrix
\begin{equation}
\label{Smatrix}
S_{\pm}=\Pi R_{\pm} \ ,
\end{equation}
determines scattering processes. Even though specific scattering amplitudes depend on $\delta_N$, every S-matrix gives rise to the same set of selection rules between `in' and `out' states.
\begin{itemize}
\item{Scattering processes are elastic: transition amplitudes vanish unless $m_{\inn}+n_{\inn}=m_{\outt}+n_{\outt}$.}
\item{Scattering processes preserve parity: transition amplitudes vanish unless $m_{\inn}$ and $m_{\outt}$ are both even or odd, and $n_{\inn}$ and $n_{\outt}$ are both even or odd.}
\item{The bosonic channel $\varphi\varphi,\psi\psi\mapsto\varphi\varphi,\psi\psi$ corresponds to $m_{\inn}-n_{\inn}=n_{\outt}-m_{\outt}\in4\mathbb{Z}$.}
\item{The fermionic channel $\varphi\psi,\psi\varphi\mapsto\varphi\psi,\psi\varphi$ corresponds to $m_{\inn}-n_{\inn}=n_{\outt}-m_{\outt}\in4\mathbb{Z}+2$.}
\end{itemize}

From now on, we will focus on the pair of R-matrices $R_{-, 1}\equiv R_{-}$ and $R_{+, 4}\equiv R_{+}$  (under $a_+\mapsto-a_+$), that is
\begin{equation}
\label{mats}
R_{+}=
\begin{bmatrix}
{d_{+}} & 0 & 0 & -{a_{+}} \\
0 & -{d_{+}} & {a_{+}} & 0 \\
0 & {a_{+}} & {d_{+}} & 0 \\
-{a_{+}} & 0 & 0 & -{d_{+}}
\end{bmatrix} 
\ , \quad R_{-}=
\begin{bmatrix}
{d_{-}} & 0 & 0 & {a_{-}} \\
0 & {d_{-}} & {a_{-}} & 0 \\
0 & {a_{-}} & -{d_{-}} & 0 \\
{a_{-}} & 0 & 0 & -{d_{-}}
\end{bmatrix}  \ ,
\end{equation}
and we will choose 
\begin{equation}
\label{da}
\begin{split}
d_{\pm}\ket{m,n}&=\overset{\infty}{\underset{N=0}{\sum}}c_N \e^{\mp N\theta}\ket{m\mp4N,n\pm4N} \ ,\\
a_{\pm}\ket{m,n}&=\overset{\infty}{\underset{N=0}{\sum}}c_N \e^{\mp N\theta\mp\theta/2}\,\ket{m\mp4N\mp2,n\pm4N\pm2}\ ,
\end{split}
\end{equation}
where $c_N$ are coefficients we will determine later. We will assume $\mathfrak{R}\theta>0$ for $d_{+}$ and $a_{+}$, and $\mathfrak{R}\theta<0$ for $d_{-}$ and $a_{-}$. These assumptions make $d_\pm$ and $a_\pm$ well-behaved linear operators on the momentum space of two particles, where they are bounded and furthermore invertible.

Equations (\ref{mats1}), (\ref{mats2}), and (\ref{alphadelta}) determine general solutions to the YB equation. Furthermore, we could have defined~$d_\pm$ and~$a_\pm$ in (\ref{daans}) with coordinated shifts on states or alternative values of $N$ in the sum, and they would have solved the YB equation too. The reason for the choice of~(\ref{mats}) is that $R_{\pm}$ are two of the simplest non-trivial choices that exhibit a set of properties we want to hold. First, each $R_{\pm}$ enjoys the right symmetries. Each R-matrix is relativistic invariant, and invariant under (an infinite-dimensional embedding of) $\alg$, the superalgebra we introduced in section \ref{salg}. Both $R_{\pm}$ are also crossing-symmetric and, together, satisfy the form of braiding unitarity introduced in \cite{Andrea2}. Finally,  the IR limit of $R_{+}$ and $R_{-}$ reproduces solution 3 and solution 5 of \cite{Andrea2}, respectively. We will devote the remaining sections to analyse these properties. We will begin with the matching between the IR limit of $R_{\pm}$ and relativistic massless R-matrices of \cite{Andrea2}, which will provide a first benchmark for the choice of (\ref{mats}).

\section{The infrared limit of the R-matrices}

\label{sads2}

In this section, we will retrieve solutions 3 and 5 of \cite{Andrea2} from $R_+$ and $R_-$, respectively. The solutions \cite{Andrea2} appear, in particular, in the IR limit of (\ref{mats}). To apply the limit, we write the spectral decomposition of (\ref{da}) first:
\begin{equation}
\label{dapreIR}
\begin{split}
d_{\pm}=\overset{\infty}{\underset{m_{\outt}=-\infty}{\sum}}\,\overset{\infty}{\underset{n_{\outt}=-\infty}{\sum}}\,\overset{\infty}{\underset{m_{\inn}=-\infty}{\sum}}\,\overset{\infty}{\underset{n_{\inn}=-\infty}{\sum}}\bra{m_{\outt},n_{\outt
}}d_\pm\ket{m_{\inn},n_{\inn}} \ket{m_{\outt},n_{\outt}}\bra{m_{\inn},n_{\inn}} \ , \\
a_{\pm}=\overset{\infty}{\underset{m_{\outt}=-\infty}{\sum}}\,\overset{\infty}{\underset{n_{\outt}=-\infty}{\sum}}\,\overset{\infty}{\underset{m_{\inn}=-\infty}{\sum}}\,\overset{\infty}{\underset{n_{\inn}=-\infty}{\sum}}\bra{m_{\outt},n_{\outt
}}a_\pm\ket{m_{\inn},n_{\inn}} \ket{m_{\outt},n_{\outt}}\bra{m_{\inn},n_{\inn}} \ ,
\end{split}
\end{equation}
where
\begin{equation}
\label{transda}
\begin{split}
\bra{m_{\outt},n_{\outt
}}d_\pm\ket{m_{\inn},n_{\inn}}=\frac{1}{(2\pi L)^2}\int_{-\pi L}^{\pi L}\dif\phi \e^{-\im\,(m_{\outt}-m_{\inn})\phi/L  } \int_{-\pi L}^{\pi L}\dif\phi^{\prime}\e^{-\im\,(n_{\outt}-n_{\inn})\phi^{\prime}/L}\\
\times 
\overset{\infty}{\underset{N=0}{\sum}}c_N\e^{\mp N(\theta+\im4\phi/L-4\im\phi^{\prime}/L)\,}\ , \\
\bra{m_{\outt},n_{\outt
}}a_\pm\ket{m_{\inn},n_{\inn}}=\frac{\e^{\mp\theta/2}}{(2\pi L)^2}\int_{-\pi L}^{\pi L}\dif\phi \e^{-\im\,(m_{\outt}-m_{\inn}\pm2) \phi/L  } \int_{-\pi L}^{\pi L}\dif\phi^{\prime}\e^{-\im\,(n_{\outt}-n_{\inn}\mp2) \phi^{\prime}/L }\\
\times 
\overset{\infty}{\underset{N=0}{\sum}}c_N\e^{\mp N(\theta+\im4\phi/L-4\im\phi^{\prime}/L)\,}\ .
\end{split}
\end{equation}
To write these formulae, we have used $\mathfrak{R}\theta>0$ for $d_{+}$ and $a_{+}$, and $\mathfrak{R}\theta<0$ for $d_{-}$ and $a_{-}$, and interchanged sums and integrals.

Solutions 3 and 5 of \cite{Andrea2} are $(4\times4)$ supermatrices that solve the YB equation. These supermatrices are blind to momentum space, hence we are bound to consider a limit of~$R_\pm$ that trivialises the action inside $\Ls\otimes\Ls$. The limit we look for is the IR limit $L\rightarrow \infty$, which decompactifies the circle $\mathrm{S}^1$ into the real line $\mathbb{R}$. To apply the limit in the momentum space, we need to introduce the one-dimensional wave vector $k_n=n/L$. The IR limit is then the standard continuum limit
\begin{equation}
\label{continuum}
\phi\in[-\pi L,\pi L)\rightarrow x \in\mathbb{R} \ , \quad k_n=\frac{n}{L}\in\frac{1}{L}\mathbb{Z}\rightarrow k\in\mathbb{R} \ ,  \quad \frac{1}{L}\overset{\infty}{\underset{n=-\infty}{\sum}}f_n\rightarrow\int_{-\infty}^{\infty}\dif k f(k) \ ,
\end{equation}
where $f\in L^2(\mathbb{R})$. 

The application of the IR limit is not enough for (\ref{dapreIR}) to provide us with solutions 3 and 5 of \cite{Andrea2}. The problem is that the IR limit alone does not trivialise the action inside the momentum space. The pair~$d_\pm$ and $a_\pm$ in (\ref{da}) allow for non-vanishing transition amplitudes between states with arbitrarily large $m_{\outt}-m_{\inn}$ and $n_{\outt}-n_{\inn}$. Therefore,~(\ref{continuum}) does not delete the contribution of the upper region in the sums of (\ref{da}), which should leave a finite remnant. To erase the contribution of large $m_{\outt}-m_{\inn}$ and $n_{\outt}-n_{\inn}$ in the sums, we need to properly apply $\theta\rightarrow\mp\infty$ to the transition amplitudes of $a_{\pm}$ and $d_{\pm}$ before the IR limit.

If we apply $\abs{\theta}\rightarrow\infty$ and the IR limit $L\rightarrow\infty$ afterwards, we can rephrase~(\ref{dapreIR}) like
\begin{equation}
\begin{split}
d_{\pm}=\int_{\infty}^{\infty}\dif k_{\outt}\int_{\infty}^{\infty}\dif k^{\prime}_{\outt}\int_{\infty}^{\infty}\dif k_{\inn}\int_{\infty}^{\infty}\dif k_{\inn}^{\prime}\bra{k_{\outt},k^{\prime}_{\outt
}}d_\pm\ket{k_{\inn},k_{\inn}^{\prime}} \ket{k_{\outt},k_{\outt}^{\prime}}\bra{k_{\inn},k_{\inn}^{\prime}} \ , \\
a_{\pm}=\int_{\infty}^{\infty}\dif k_{\outt}\int_{\infty}^{\infty}\dif k^{\prime}_{\outt}\int_{\infty}^{\infty}\dif k_{\inn}\int_{\infty}^{\infty}\dif k_{\inn}^{\prime}\bra{k_{\outt},k^{\prime}_{\outt
}}a_\pm\ket{k_{\inn},k_{\inn}^{\prime}} \ket{k_{\outt},k_{\outt}^{\prime}}\bra{k_{\inn},k_{\inn}^{\prime}}   \ ,
\end{split}
\end{equation}
where
\begin{equation}
\begin{split}
\bra{k_{\outt},k^{\prime}_{\outt
}}d_\pm\ket{k_{\inn},k_{\inn}^{\prime}}=\delta(k_{\outt}-k_{\inn})\delta(k^{\prime}_{\outt}-k^{\prime}_{\inn})\left[c_0+\mathrm{O}(\e^{\mp\theta})\right] \ , \\
\bra{k_{\outt},k^{\prime}_{\outt
}}a_\pm\ket{k_{\inn},k_{\inn}^{\prime}}=\e^{\mp\theta/2}\delta(k_{\outt}-k_{\inn})\delta(k^{\prime}_{\outt}-k^{\prime}_{\inn})\left[c_0+\mathrm{O}(\e^{\mp\theta})\right] \ , \\
\end{split}
\end{equation}
and we have introduced the (generalised) orthonormal basis of plane waves of $L^2(\mathbb{R})$:
\begin{equation}
\left\{\ket{k}:\bra{x}\ket{k}=\frac{1}{\sqrt{2\pi}}\e^{\im k x} \ , \ k\in\mathbb{R}\right\} \ .
\end{equation}
Finally, we need to average $d_\pm$ and $a_\pm$, hence $R_\pm$, over the momenta of particles by the partial trace. This step is necessary to obtain fully fledged $(4\times4)$ supermatrices. The computation of the partial trace actually demands the regularisation of the result to account for divergence of the volume of the momentum space of two particles and two $\delta$-functions evaluated at zero. We regularise the result by the partial trace of the identity over of the momentum space of two particles, namely
\begin{equation}
\label{N}
\mathcal{N}=\int_{-\infty}^{\infty}\dif k \int_{-\infty}^{\infty}\dif k^{\prime}\bra{k,k^{\prime}}\ket{k,k^{\prime}} \ .
\end{equation}
If we compute the partial trace of $R_{\pm}$ in this way, we obtain
\begin{equation}
\label{R3}
R_3= \frac{1}{\mathcal{N}}\int_{-\infty}^{\infty}\dif k\int_{-\infty}^{\infty}\dif k^{\prime} \bra{k,k^{\prime}}R_+\ket{k,k^{\prime}}  
=c_0\begin{bmatrix}
1 & 0 & 0 & -\e^{-\theta/2}\\
0 & -1 & \e^{-\theta/2}\ & 0 \\
0 & \e^{-\theta/2}\ & 1 & 0 \\
-\e^{-\theta/2}\ & 0 & 0 & -1
\end{bmatrix} \ ,
\end{equation}
up to terms $\mathrm{O}(\e^{-\theta})$ \ , and
\begin{equation}
\label{R5}
R_5= \frac{1}{\mathcal{N}}\int_{-\infty}^{\infty}\dif k\int_{-\infty}^{\infty}\dif k^{\prime} \bra{k,k^{\prime}}R_-\ket{k,k^{\prime}}=c_0
\begin{bmatrix}
1 & 0 & 0 & \e^{\theta/2} \\
0 & 1 & \e^{\theta/2} & 0 \\
0 & \e^{\theta/2} & -1 & 0 \\
\e^{\theta/2} & 0 & 0 & -1
\end{bmatrix}   \ ,
\end{equation}
up to terms $\mathrm{O}(\e^{\theta})$. If we set 
\begin{equation}
\label{c0}
c_0=1 \ ,    
\end{equation}
equations (\ref{R3}) and (\ref{R5}) match solution 3 (2.19) and solution 5 (2.21) of \cite{Andrea2}, respectively (under the choice of the right representation and $\alpha=1$, see footnote~\ref{fn}).
We will assume the choice (\ref{c0}) in the definition (\ref{da}) of $d_\pm$ and $a_\pm$ hereafter.

We close this section by noting that we can understand $\abs{\theta}\rightarrow\infty$ as part of the IR limit. From this point of view, $\theta$ sets a scale for the scattering that the two-body S-matrices~$S_\pm$ defined in~(\ref{Smatrix}) determines. If $\abs{\theta}$ is small, both particles carry almost the same rapidity. The band of allowed transitions between integer incoming momenta $m_{\inn}$ and $n_{\inn}$, and integer outgoing momenta $m_{\outt}$ and $n_{\outt}$ is arbitrarily wide. Scattering occurs in the `ultraviolet' regime, which involves the momentum space as a whole. On the contrary, if $\abs{\theta}\rightarrow\infty$ is large, the relativistic rapidities of both particles are largely different. Scattering just occurs between particles with almost the same $m_{\inn}$ and $n_{\inn}$, and $m_{\outt}$ and $n_{\outt}$. The IR limit $L\rightarrow\infty$ just reinforces this picture by imposing that the incoming momenta $k_{\inn}$ and $k_{\inn}^{\prime}$ (which are no longer integers) must equal the outgoing momenta $k_{\outt}$ and $k^{\prime}_{\outt}$ for particles to scatter. This picture compels us to consider $\abs{\theta}\rightarrow\infty$ and $L\rightarrow\infty$ as parts of the same IR limit. Moreover, the picture also makes clear why the average over the momenta of particles still provides an accurate representation of scattering. Since the scattering in the momentum space is trivial, the partial trace is just the coarse-grain sieve that we need to obtain the effective description of the scattering of massless modes in the IR limit.

\section{Braiding unitarity}

\label{bus}

Having proved that the R-matrices $R_{\pm}$ reproduce the R-matrices of \cite{Andrea2}, we will show, in this section, that we can fix the coefficients $c_N$ of $d_{\pm}$ and $a_{\pm}$ in a such way that $R_{\pm}$ fulfil modified braiding unitarity as proposed in \cite{Andrea2}. Reference \cite{Andrea2}, specifically, demonstrated that solution 3 (\ref{R3}) and solution 5 (\ref{R5}) satisfy
\begin{equation}
\label{mbu}
    R_3(\theta) R_5^{\opp} (-\theta)= 1\otimes 1 \ .
\end{equation}
Then,  \cite{Andrea2} argued that (\ref{mbu}) is the general condition of modified brading unitarity, in the sense that (\ref{mbu}) appears under the application of $\theta\rightarrow\infty$ to the braiding unitarity condition of a given R-matrix.

Since $R_\pm$ match solution 3 and solution 5 of \cite{Andrea2}, we demand these R-matrices to satisfy modified braiding unitary:
\begin{eqnarray}
\label{mbuinf}
R_+(\theta) \, R_-^{\opp}(-\theta) = 1\otimes 1 \ ,
\end{eqnarray}
where $R_-^{\opp}=\Pi  R_-  \Pi$. The definition of $\Pi$ implies that $R_-^{\opp}$ is just $R_+$ in (\ref{mats}) under the replacement of $d_+$ and $a_+$ by
\begin{equation}
    \Pi  d_{-} (-\theta)  \Pi= d_{+} (\theta) \ , \qquad \Pi  a_{-} (-\theta)  \Pi= a_{+} (\theta) \ ,
\end{equation}
respectively. Therefore, (\ref{mbuinf}) is equivalent to impose that~$R_+$ (and, consequently, $R_-$) is self-inverse. Hence, (\ref{mbuinf}) implies two infinite sets of equations:
\begin{equation}
\label{invcond}
(d_{+}d_{+}+a_{+}a_{+})\ket{m,n}=\ket{m,n} \ , \quad
(d_{+}a_{+}-a_{+}d_{+})\ket{m,n}=0\ ,
\end{equation}
where $\ket{m,n}$ is arbitrary. If we use (\ref{da}), we deduce that the second set is trivial. On the other hand, if we employ (\ref{da}) in the first set, we deduce an infinite system of quadratic equations on the coefficients $c_{N}$, namely
\begin{align}
\label{cc}
\sum_{n=0}^N c_n c_{N-n}=(-1)^N ,
\end{align}
for every $N>0$. To solve (\ref{cc}), we rely on the method of the generating function.  Let
us introduce
\begin{equation}
g(x)=\sum_{N=0}^\infty c_N x^N \ .
\end{equation}
Therefore,
\begin{equation}
g^2(x)=\sum_{N=0}^\infty \sum_{n=0}^N c_n c_{N-n} x^N = \sum_{N=0}^\infty (-1)^N x^N =\frac{1}{1+x} \ .
\end{equation}
If we take the positive branch, we obtain,
\begin{equation}
\label{cN}
c_N = \frac{(-1)^N \, (2N)!}{4^N \, (N!)^2} \ ,
\end{equation}
which, in particular, provides (\ref{c0}) at $N=0$. If we pick these $c_N$, we will obtain that modified braiding unitarity (\ref{mbuinf}) between $R_{\pm}$ holds. We will assume (\ref{cN}) in the following.

If we had multiplied $R_{\pm}$ by any factor $\Omega_\pm=\exp(\Phi_\pm)$ , where~$\Phi_\pm$ denotes the dressing phase (that, in principle, depends on $\theta$), the product would have still solved (\ref{YB}). However, $\Omega_\pm$ must fulfil
\begin{eqnarray}
\Omega_+(\theta)\Omega_-(-\theta) = 1 \ ,
\end{eqnarray} 
for modified braiding unitary (\ref{mbuinf}) to hold.

\section{Symmetries of the R-matrices}

\label{ssym}

In this section, we analyse the symmetry superalgebra of R-matrices $R_{\pm}$. We will show that we need an infinite-dimensional embedding of $\alg$, the superalgebra we introduced in section~\ref{salg}, to construct the symmetry superalgebra of $R_{\pm}$. We will also use the Lorentz boost to prove that $R_{\pm}$ is relativistic invariant. 

By definition, the R-matrices $R_\pm$ are invariant under (a representation of) the superalgebra $\algf$ with respect to the coproduct $\Delta$ if $\algf$ admits the homomorphism $\Delta:\algf\rightarrow\algf\otimes\algf$ such that
\begin{eqnarray}
\Delta^{\opp}(\Xg) R_\pm = R_\pm \Delta(\Xg) \ , \label{copo}
\end{eqnarray}
for every $\Xg\in\algf$, where $\Delta^{\opp} = \Pi \circ \Delta$. The R-matrices $R_\pm$ are invariant under $\alg$ with respect to the cocommutative coproduct 
\begin{align}
\label{coproduct}
&\Delta(\Qg) = \Qg \otimes \Dg^2 + \Dg^2 \otimes \Qg \ , & &\Delta(\Pg) = \Pg \otimes \Dg^4 + \Dg^4 \otimes \Pg \ , & &\Delta(\Hg) = \Hg \otimes \Dg^2 + \Dg^2 \otimes \Hg  \ ,  \notag\\
 &\Delta(\Sg) = \Sg \otimes 1 + 1 \otimes \Sg \ , & &\Delta(\Kg) = \Kg \otimes 1 + 1 \otimes \Kg \ , & &\Delta(\Cg) = \Cg \otimes 1 + 1 \otimes \Cg \ ,   \\
&\Delta(\Dg) =\Dg \otimes \Dg  \ , & &\Delta(\Dg^{-1}) =\Dg^{-1} \otimes \Dg^{-1}  \ . \notag
\end{align}
The coproduct above is not a homomorphism of the superalgebra $\alg$ itself. The obstruction is posed by the commutators between $\Cg$ and $\Qg$, $\Pg$, $\Hg$, $\Dg$ and $\Dg^{-1}$. Instead, $\Delta$ is a homomorphism of $\algl$, an infinite-dimensional superalgebra that embeds $\alg$, hence $\algl$ rather than $\alg$ is the symmetry superalgebra of ${R}_\pm$. Let us be more precise about $\algl$. 

We define $\algl$ to be an infinite-dimensional loop-like superalgebra where $\alg$ plays the role of zeroth level. We base $\algl$ on the primitive set $\Qg_A$, $\Sg_A$, $\Dg_A$, $\Dg_{A}^{-1}$, and $\Cg_0$ (together with the unity 1) where $A=0,1,2,\dots$ denotes the level. (We do not need to introduce the tower of fermions $\Sg_A$ to obtain the symmetry superalgebra of $R_{\pm}$, but we do so for $\Qg_{A}$ and $\Sg_{A}$ to appear on equal footing.) The boson $\Cg_0$ shifts the level inside $\algl$:
\begin{equation}
 \begin{split}
[\Cg_0 , \Qg_A] &= -2 \Qg_{A+1} \ ,  \quad [\Cg_0, \Sg_A] = 2 \Sg_{A+1} \ , \\
\quad [\Cg_0, \Dg_{A}] &=-2 \Dg_{A+1} \ , \quad [\Cg_0, \Dg_{A}^{-1}] =2 \Dg_{A+1}^{-1} \ .
\end{split}
\end{equation}
The towers of bosons $\Dg_{A}$ and $\Dg_{A}^{-1}$ commutes with every other generator but $\Cg_0$. Finally, the anticommutation relations between the towers of fermions $\Qg_A$ and $\Sg_A$ generate the set of composite bosons $\Pg_{A,B}$, $\Kg_{A,B}$, and $\Hg_{A,B}$:
\begin{equation}
\label{PKH}
\{\Qg_{A},\Qg_{B}\}=\Pg_{A,B} \ , \quad \{\Sg_{A},\Sg_{B}\}=\Kg_{A,B} \ , \quad \{\Qg_{A},\Sg_{B}\}=\Hg_{A,B} \ . 
\end{equation}
We could consistently obtain more generators in $\algl$ by the iteration of the (anti-)commutation relations. We will not be concerned with the minimal number of generators we need to span $\algl$ linearly. We will consider the primitive set of $\Qg_A$, $\Sg_A$, $\Dg_A$, and $\Cg_0$ instead. The coproduct
\begin{equation}
\begin{split}
    \Delta(\Qg_A) &= \sum_{B=0}^A \sum_{C=0}^{A-B} \binom{A}{B,C,A-B-C}  ( \Qg_B \otimes \Dg_C \Dg_{A-B-C} + \Dg_C \Dg_{A-B-C} \otimes \Qg_B ) \ ,  \\
    \Delta(\Sg_A) &= \Sg_A \otimes 1 + 1 \otimes \Sg_A \ ,  \\
    \Delta(\Dg_A) &= \sum_{B=0}^A \binom{A}{B} \Big( \Dg_B \otimes \Dg_{A-B} + \Dg_{A-B} \otimes \Dg_B \Big) \ ,  \\
    \Delta(\Dg_A^{-1}) &=\sum_{B=0}^A \binom{A}{B} ( \Dg_B^{-1} \otimes \Dg_{A-B}^{-1} + \Dg_{A-B}^{-1} \otimes \Dg_B^{-1} )  \ ,  \\
    \Delta(\Cg_0) &= \Cg_0 \otimes 1 + 1 \otimes \Cg_0 \ , \label{gencoproduct}
\end{split}
\end{equation}
where we have introduced binomial and multinomial coefficients, is a homomorphism of $\algl$. The coproduct of composite generators inside $\algl$, for instance, $\Hg_{A,B}$ follows from the fact $\Delta$ is a homomorphism. We now introduce the representation over $\mathcal{H}$, 
\begin{equation}
\label{repinf}
\Qg_{A}=\Qg \ , \quad \Sg_{A}=\Sg \ , \quad \Dg_{A}=\Dg \ , \quad \Dg_{A}^{-1}=\Dg^{-1} \ , \quad \Cg_{0}=\Cg \ ,    
\end{equation}
where $\Qg$, $\Sg$, $\Dg$, $\Dg^{-1}$, and $\Cg$ are defined in (\ref{repmom}) and (\ref{repD}). In the representation (\ref{repinf}) $\algl$ closes because $\alg$ closes. The coproduct (\ref{gencoproduct}) reduces to
\begin{equation}
\begin{split}
\label{coproductinf}
&\Delta(\Qg_A) =3^{A}( \Qg \otimes \Dg^2 + \Dg^2 \otimes \Qg ) \ , \quad \Delta(\Sg_A) = \Sg \otimes 1 + 1 \otimes \Sg  , \\  
&\Delta(\Cg_0) = \Cg \otimes 1 + 1 \otimes \Cg \ , \quad \Delta(\Dg_A) =2^{A}(\Dg \otimes \Dg)  \ , \quad \Delta(\Dg^{-1}) =2^{A}(\Dg^{-1} \otimes \Dg^{-1})  \ .
\end{split}
\end{equation}
Therefore, the coproduct $\Delta$ we introduced in (\ref{coproduct}) is nothing but the coproduct of $\algl$ at zeroth level, that is, at the level of $\alg$. We deduce that $R_{\pm}$ is invariant under the representation (\ref{repinf}) of $\algl$ with respect to the coproduct (\ref{gencoproduct}); in other words, $\algl$ is the symmetry superalgebra of $R_{\pm}$.

Up to this point, we have set aside the Lorentz boost $\Jg$. We can promote the action of $\Jg$ to an outer automorphism on $\algl$ into itself by the straightforward generalisation of (\ref{Jcomm}), (\ref{JC}), and (\ref{JD}):
\begin{equation}
\label{Jcomminf}
[\Jg , \Qg_A]=\frac{1}{2} \Qg_A \ , \quad [\Jg , \Sg_A]=\frac{1}{2} \Sg_A \ , \quad [\Jg , \Dg_A]=0 \ , \quad [\Jg , \Dg_A^{-1}]=0 \ , \quad [\Jg , \Cg]= 0 \ .
\end{equation}
The trivial coproduct
\begin{equation}
\label{copJ}
\Delta(\Jg) = \Jg \otimes 1 + 1 \otimes \Jg \ ,
\end{equation}
is an enhancement of (\ref{gencoproduct}) compatible with (\ref{Jcomminf}). If we use (\ref{Jrep}), the representation of $\Jg$ over $\mathcal{H}$, we deduce that $R_\pm$ are invariant under $\Jg$ with respect to (\ref{copJ}). The invariance under $\Jg$ is just the customary relativistic invariance of the S-matrices $S_{\pm}$ associated to $R_{\pm}$, which depend on the relativistic rapidities $\theta_a$ only through $\theta=\theta_1 - \theta_2$.

We close the section by noting that (\ref{coproduct}) reduces to the trivial coproduct under the application of the IR limit in the momentum space of two particles we detailed in section \ref{sads2}. The reduction follows from the trivialisation of $\Dg$, $\Dg^{-1}$, and $\Cg$ in the IR limit, whose action we can neglect once we apply the coarse-grain sieve over the momentum space. The trivialisation of (\ref{coproduct}) is consistent with the fact $R_3$ in (\ref{R3}) and $R_5$ in (\ref{R5})  are invariant under the relativistic limit of massless representation of $\psuR$ with respect to the trivial coproduct \cite{Andrea2}.

\section{Invariance of the R-matrices under crossing symmetry}

\label{sRcross}

In this section, we prove that the R-matrices $R_\pm$ are invariant under crossing symmetry. In subsection \ref{ars}, we introduce the antiparticle representation of $\alg$ and $\algl$. In subsection \ref{css}, we use the antiparticle representation to obtain crossed R-matrices and prove that $R_\pm$ are crossing symmetric.

\subsection{Antiparticle representation}

\label{ars}

To construct the antiparticle representation of a given superalgebra $\algf$, we must promote $\algf$ to a Hopf algebra. The Hopf algebra consists of $\algf$ endowed with the product $\mu:\algf\otimes\algf\rightarrow\algf$, the coproduct $\Delta:\algf\rightarrow\algf\otimes\algf$, the unit $\eta:\mathbb{C}\rightarrow\algf$, the counit $\epsilon:\algf\rightarrow\mathbb{C}$ and the antipode $s:\algf\rightarrow\algf$, subject to some compatibility conditions. We cannot promote $\alg$ to a Hopf algebra because $\Delta$ in (\ref{coproduct}) is not a homomorphism of $\alg$. Nonetheless, we can promote the infinite-dimensional embedding~$\algl$ to a Hopf algebra. Let us detail the way to enhance $\algl$. (For brevity, we will not update the action of $\Jg$ on $\algl$ to the Hopf algebra; the relativistic invariance of crossed R-matrices is reminiscent from the hidden presence of $\Jg$ in the Hopf algebra).

First, we identify the product $\mu$ with the multiplication of elements of $\algl$ and the coproduct~$\Delta$ with (\ref{gencoproduct}). For $\algl$ to be a Hopf algebra, $\mu$ must be associative and $\Delta$ must be coassociative. Both properties hold. The product must be also compatible with the unit:
\begin{equation}
\mu (\eta\otimes1) (x\otimes \Xg) = \mu (1\otimes\eta) (\Xg\otimes x)=x\Xg \ ,
\end{equation}
for all $x\in\mathbb{C}$ and $\Xg\in\algl$. The unit reads
\begin{equation}
\eta(x)=x1 \ .
\end{equation}
Moreover, the coproduct $\Delta$ must be compatible with the counit $\epsilon$:
\begin{equation}
(\epsilon \otimes 1) \Delta(\Xg) = (1 \otimes \epsilon)  \Delta(\Xg)= \Xg  \ , \label{fundo}
\end{equation}
for all $\Xg\in\algl$. It suffices to specify $\epsilon$ on the primitive set $1$, $\Qg_A$, $\Sg_A$, $\Dg_A$, $\Dg_A^{-1}$, and $\Cg_0$. The action of $\epsilon$ on the remaining elements of $\algl$ follows from the fact that $\Delta$ in (\ref{gencoproduct}) is a homomorphism. The counit reads
\begin{equation}
\epsilon(1)=1 \ , \quad \epsilon(\Qg_{A})=0 \ , \quad  \epsilon(\Sg_{A})=0 \ , \quad \epsilon(\Dg_A)=\delta_{A}^{0} \ , \quad \epsilon(\Dg_A^{-1})=\delta_{A}^{0} \ , \quad \epsilon(\Cg_{0})=0 \ .
\end{equation}
Finally, the antipode $s$ must be compatible with the fact $\algl$ is an algebra and coalgebra
\begin{equation}
\mu (s \otimes 1) \Delta (\Xg) = \mu (1 \otimes s) \Delta (\Xg) = \eta (\epsilon(\Xg)) \ .  \label{fund}
\end{equation}
We will only specify the action of $s$ on $1$, $\Qg_0$, $\Sg_0$, $\Dg_0$, $\Dg_0^{-1}$, and $\Cg_0$. The action of $s$ on both the remaining primitive and composite elements of $\algl$ follows from the fact~$\Delta$ in (\ref{gencoproduct}) is a homomorphism and $s$ an antihomomorphism. The reason to omit the action of $s$ on $\Qg_A$, $\Sg_A$, $\Dg_A$, $\Dg_A^{-1}$ with $A=1,2\dots$ is that we have not been able to derive closed formulae for them, although its iterative computation is straightforward. The antipode reads
\begin{equation}
\label{antipode}
s(1)=1 \ , \quad s(\Qg_{0})=-(\Dg_0^{-1})^4\Qg_0\ , \quad  s(\Sg_{0})=-\Sg_0 \ , \quad s(\Dg_0)=\Dg_0^{-1} \ , \quad s(\Dg_0^{-1})=\Dg_0 \ .
\end{equation}
We emphasise that (\ref{antipode}) makes manifest that the presence of $\Dg_0^{-1}$ is necessary inside $\algl$, otherwise we had not been able to enhance $\algl$ with the antipode.

Once we have promoted $\algl$ to a Hopf algebra, we can construct the antiparticle representation of (\ref{repinf}), in turn based on (\ref{repmom}) and (\ref{repD}), over $\mathcal{H}$. By definition, the antiparticle $\bar{\Xg}$ of the particle $\Xg$ is
\begin{equation}
    s(\Xg(\theta)) = C^{-1} \bar{\Xg}^{\mathrm{st}} (\theta+\im\pi) C \ ,
\end{equation}
where $C$ denotes the charge conjugation 
\begin{equation}
C=
\begin{bmatrix}
    \im & 0 \\
    0 & 1 
 \end{bmatrix} \ ,
\end{equation}
and, if
\begin{equation}
M=
 \begin{bmatrix}
    M_{1}^{1} & M_{2}^{1} \\
    M^{2}_{1} & M^{2}_{2}
\end{bmatrix} \ , \quad 
M_{\beta}^{\alpha}\ket{m}=\overset{\infty}{\underset{n=-\infty}{\sum}}\mu^{\alpha,m}_{\beta,n}\ket{m-n} \ ,
\end{equation}
the definition of supertranspose is 
\begin{equation}
M^{\mathrm{st}}=
\begin{bmatrix}
    \left(M_{1}^{1}\right)^{\mathrm{st}} & \left(M^{2}_{1}\right)^{\mathrm{st}} \\
    -\left(M^{1}_{2}\right)^{\mathrm{st}} & \left(M^{2}_{2}\right)^{\mathrm{st}}
\end{bmatrix} \ , \quad (M_{\beta}^{\alpha})^{\mathrm{st}} \ket{m}=\overset{\infty}{\underset{n=-\infty}{\sum}}\mu^{\alpha,m}_{\beta,n}\ket{m+n} \ .
\end{equation}
The antiparticles of the particles (\ref{repmom}) of $\alg$  are
\begin{alignat}{4}
\label{repmomantip}
    \bar{\Qg}=\frac{1}{\sqrt{2}}\e^{\theta/2+\im\pi/4} 
    \begin{bmatrix} 
    0 & (b^{\dagger})^3 \\ 
    (b^{\dagger})^3 & 0  
    \end{bmatrix} &\ , \quad&
    \bar{\Pg}=\frac{\im}{2}\e^{\theta} 
    \begin{bmatrix} 
    (b^{\dagger})^6 & 0 \\
    0 & (b^{\dagger})^6 
    \end{bmatrix}
     &\ ,  \quad&
    \bar{\Hg}=\frac{1}{2}\e^{\theta} 
    \begin{bmatrix} 
    (b^{\dagger})^4 & 0 \\
    0 & (b^{\dagger})^4  
    \end{bmatrix} &\ ,& \nonumber \\
    \bar{\Sg}=\frac{1}{\sqrt{2}}\e^{\theta/2-\im\pi/4} 
    \begin{bmatrix} 
    0 & b^{\dagger} \\
    b^{\dagger} & 0  
    \end{bmatrix} &\ , \quad&
    \bar{\Kg}=-\frac{\im}{2}\e^{\theta} 
    \begin{bmatrix} 
    (b^{\dagger})^2 & 0 \\
    0 & (b^{\dagger})^2 
    \end{bmatrix} &\ , \quad&
    \bar{\Cg}=-
    \begin{bmatrix} 
    N & 0 \\
    0 & N  
    \end{bmatrix} &\ ,&
\end{alignat}
whereas $\bar{\Dg}=\Dg$ and $\bar{\Dg}^{-1}=\Dg^{-1}$. The antiparticle representation of $\algl$ straightforwardly follows from (\ref{repinf}).

\subsection{Crossed R-matrices and crossing symmetry}
 \label{css}

To prove crossing symmetry, we have to construct the crossed R-matrices of $R_{\pm}$, the R-matrix of the particle-particle channel: the R-matrices $R_\pm^{\mathrm{p\bar{p}}}$ of the particle-antiparticle channel, $R_\pm^{\mathrm{\bar{p}p}}$ of the antiparticle-particle channel, and $R_\pm^{\mathrm{\bar{p}\bar{p}}}$ of the antiparticle-antiparticle channel. The construction of crossed R-matrices is based on the antiparticle representation (\ref{repmomantip}). By definition, the crossed R-matrices of $R_\pm$ fulfil
\begin{equation}
\begin{split}
(\Delta^{\opp})^{\mathrm{\bar{p}p}}(\Xg) R_\pm^{\mathrm{\bar{p}p}}= R_\pm^{\mathrm{\bar{p}p}} \Delta^{\mathrm{\bar{p}p}}(\Xg)  \ , \\
(\Delta^{\opp})^{\mathrm{p\bar{p}}}(\Xg) R_\pm^{\mathrm{p\bar{p}}}= R_\pm^{\mathrm{p\bar{p}}} \Delta^{\mathrm{p\bar{p}}}(\Xg) \ , \\
(\Delta^{\opp})^{\mathrm{\bar{p}\bar{p}}}(\Xg) R_\pm^{\mathrm{\bar{p}\bar{p}}}= R_\pm^{\mathrm{\bar{p}\bar{p}}} \Delta^{\mathrm{\bar{p}\bar{p}}}(\Xg) \ , \\
\end{split}
\end{equation}
for every $\Xg$ in $\algl$, where $\Delta^{\mathrm{\bar{p}p}}(\Xg)$, $\Delta^{\mathrm{p\bar{p}}}(\Xg)$, and $\Delta^{\mathrm{\bar{p}\bar{p}}}(\Xg)$ denote the coproducts of $\Xg$ where the first particle, the second, or both have been replaced by antiparticles, respectively.

The computation of crossed R-matrices is straightforward. The R-matrices $R_\pm^{\mathrm{\bar{p}p}}$ are just the supermatrices~(\ref{mats}) under the replacement of $d_{\pm}$ and by $a_{\pm}$ by
\begin{equation}
\begin{split}
\label{bpp}
&d_\pm^{\mathrm{\bar{p}p}}|m,n\rangle = \sum_{N=0}^\infty c_N \e^{\mp N\theta} |m\pm 4N,n\pm 4N\rangle \ , \\
&a_\pm^{\mathrm{\bar{p}p}}|m,n\rangle = \sum_{N=0}^\infty c_N \e^{\mp N\theta \mp\theta/2} |m\pm 4N \pm 2,n\pm 4N \pm 2 \rangle \ .
\end{split}
\end{equation}
The R-matrices $R_\pm^{\mathrm{p\bar{p}}}$ are the supermatrices (\ref{mats}) under the replacement of $d_{\pm}$ and by $a_{\pm}$ by
\begin{equation}
\begin{split}
\label{pbp}
&d_\pm^{\mathrm{p\bar{p}}}|m,n\rangle = \sum_{N=0}^\infty c_N \e^{\mp N\theta} |m\mp 4N,n\mp 4N\rangle \ , \\
&a_\pm^{\mathrm{p\bar{p}}}|m,n\rangle = \sum_{N=0}^\infty c_N \e^{\mp N\theta \mp \theta/2} |m\mp 4N \mp 2,n\mp 4N \mp 2\rangle \ .
\end{split}
\end{equation}
Finally, the R-matrices $R_\pm^{\mathrm{\bar{p}\bar{p}}}$ are again the supermatrices (\ref{mats})
under the replacement of $d_{\pm}$ and by $a_{\pm}$ by
\begin{equation}
\begin{split}
\label{bpbp}
&{d}^{\mathrm{\bar{p}\bar{p}}}_\pm|m,n\rangle = \sum_{N=0}^\infty c_N \e^{\mp N\theta} |m\pm 4N,n\mp 4N\rangle \ , \\
&{a}^{{\mathrm{\bar{p}\bar{p}}}}_\pm|m,n\rangle = \sum_{N=0}^\infty c_N \e^{\mp N\theta \mp \theta/2} |m\pm 4N \pm 2,n\mp 4N \mp 2 \rangle \ .
\end{split}
\end{equation}

Invariance under crossing symmetry along the particle-particle and particle-antiparticle channels imply
\begin{eqnarray}
\label{c1}
(C^{-1} \otimes 1) \, \big[R^{\mathrm{\bar{p}p}}_\pm \big]^{\mathrm{st}_1}(\theta + \im\pi) \, \, (C \otimes 1) \, R_\pm (\theta) = 1 \otimes 1 \ , \label{crosso}
\end{eqnarray} 
where $\null^{\mathrm{st}_1}$ denotes the supertranspose over the first super-Hilbert space of the graded tensor product. If we employ (\ref{mats}), (\ref{da}), and (\ref{bpp}), together with the coefficients (\ref{cN}), we straightforwardly obtain that (\ref{c1}) holds, hence~$R_{\pm}$ are invariant under crossing symmetry (along the particle-antiparticle channel). Analogously, it is straightforward to check
\begin{equation}
\begin{split}
&(1\otimes C^{-1}) \, \big[R^{\mathrm{p \bar{p}}}_\pm\big]^{\mathrm{st}_2}(\theta - \im\pi) \, \, (1 \otimes C) \, R^{\mathrm{pp}}_\pm(\theta) \, = \, 1 \otimes 1 \ ,\\
&(1\otimes C^{-1}) \, \big[R^{\mathrm{\bar{p} \bar{p}}}_\pm\big]^{\mathrm{st}_2}(\theta -\im\pi) \, \, (1 \otimes C) \, R^{\mathrm{\bar{p} p}}_\pm(\theta) \, = \, 1 \otimes 1 \ ,\\
&(C^{-1} \otimes 1) \, \big[R^{\mathrm{\bar{p} \bar{p}}}_\pm\big]^{\mathrm{st}_1}(\theta + \im\pi) \, \, (C \otimes 1) \, R^{\mathrm{p \bar{p}}}_\pm(\theta) \, = \, 1 \otimes 1 \ ,\label{crossr}
\end{split}
\end{equation}
holds, which implies crossing symmetry of $R_{\pm}$ along the corresponding channels.

Crossing symmetry (\ref{c1}) bounds $\Phi_\pm$ and~$\Phi_\pm^{\mathrm{p\bar{p}}}$ to satisfy
\begin{eqnarray}
\Omega^{\mathrm{\bar{p}p}}_\pm (\theta+\im\pi)\Omega_\pm (\theta)=1 \ ,
\end{eqnarray} 
where $\Omega^{\mathrm{\bar{p}p}}_\pm =\exp\normalsize(\Phi^{\mathrm{\bar{p} p}}_\pm \normalsize)$, and $\Phi^{\mathrm{\bar{p}p}}_\pm $ denotes the dressing phase of $R_\pm^{\mathrm{\bar{p}p}}$. Analogously,~(\ref{crossr}) implies
\begin{eqnarray}
\Omega^{\mathrm{p\bar{p}}}_\pm(\theta- \im\pi)\Omega^{\mathrm{pp}}_\pm(\theta) = 1, \qquad \Omega^{\mathrm{\bar{p}\bar{p}}}_\pm(\theta- \im\pi)\Omega^{\mathrm{\bar{p}p}}_\pm(\theta) = 1, \qquad \Omega^{\mathrm{\bar{p}\bar{p}}}_\pm(\theta+ \im\pi)\Omega^{\mathrm{p\bar{p}}}_\pm(\theta) = 1 \ ,  
\end{eqnarray}
where $\Omega^{\mathrm{p\bar{p}}}_\pm=\exp\normalsize(\Phi^{\mathrm{p\bar{p}}}_\pm \normalsize)$ and $\Omega^{\mathrm{\bar{p}\bar{p}}}_\pm =\exp\normalsize(\Phi^{\mathrm{\bar{p}\bar{p}}}_\pm\normalsize)$, and $\Phi^{\mathrm{p\bar{p}}}_\pm$ and $\Phi^{\mathrm{\bar{p}\bar{p}}}_\pm$ denote the dressing phases of $R_\pm^{\mathrm{p\bar{p}}}$ and $R^{\mathrm{\bar{p} \bar{p}}}_\pm$, respectively.

A final observation is in order. The application of the IR limit of section \ref{sads2} to (\ref{c1}) and~(\ref{crossr}) by the replacement of $R_{\pm}$ by (\ref{R3}) and (\ref{R5}) does not lead us to (3.9) in \cite{Andrea2}, the condition of invariance under crossing symmetry of $R_3$ and $R_5$. The only point at which both sets of equations match is $\abs{\theta}=\infty$. While this fact is consistent with the retrieval of $R_3$ and $R_5$ in the IR limit, which explicitly consists of~\mbox{$\abs{\theta}\rightarrow\infty$}, there is a route we can follow to overcome the general mismatch. If we were to apply the IR limit to (\ref{c1}) and (\ref{crossr}) in the same way, we would find inconsistencies. Indeed, the left-hand side requires two factors of $\mathcal{N}$ in (\ref{N}) for the regularisation to be finite, whereas the right-hand side just needs one $\mathcal{N}$ to be finite. Therefore, to apply the IR limit consistently, we need to take a second quotient by $\mathcal{N}$ in the right-hand side. We can fix the intrinsic ambiguity of this second regularisation, namely $\mathcal{N}\mapsto c(\theta)\mathcal{N}$ with $c(\theta)$ is a finite constant, by demanding compatibility with (3.9) of \cite{Andrea2} on an ad hoc basis. We could follow the same steps to obtain the modified braiding unitary conditions (\ref{mbu}) from the IR limit of (\ref{mbuinf}), although in this case the straightforward substitution of $R_{\pm}$ by $R_3$ and $R_5$ yields the same answer.

\section{Conclusions}

\label{sconclusions}

In this article, we have constructed R-matrices that generalise the relativistic limit of the scattering of massless modes with the same chirality on $\mathrm{AdS}_2$ to infinite dimensions. Our starting point has been the enhancement of $\psuR$ to $\alg$, where $\Cg$ has led us to introduce a representation of $\alg$ over the infinite-dimensional Hilbert space. Based on heuristics around the KT formula of $\gl$, we have constructed a set of infinite-dimensional R-matrices, and picked $R_{\pm}$ among them. We have defined an IR limit that enabled us to retrieve solution 3 and solution 5 of \cite{Andrea2} from $R_{+}$ and $R_{-}$, respectively. We have used modified braiding unitarity in~(\ref{mbuinf}) to determine the free coefficients of $R_{\pm}$ \cite{Andrea2}. Moreover, we proved that $R_\pm$ are invariant under $\alg$ with respect to the coproduct (\ref{coproduct}) insofar $\alg$ is embedded into the infinite-dimensional superalgebra $\algl$ in the representation (\ref{repinf}). We have considered $\algl$ as a Hopf algebra, whose antipode has permitted us to demonstrate that $R_\pm$ are crossing symmetric. Our results hint at multiple lines of research that we comment on below.

First, one may seek infinite-dimensional R-matrices beyond the relavistic limit. We have been concerned with the infinite-dimensional generalisation of the relativistic limit of the S-matrix of massless modes near the BMN vacuum  of AdS$_2$ \cite{Andrea2}. However, it seems that the non-relativistic S-matrix of massless modes written in  \cite{Hoare:2014kma} is also amenable to a parallel infinite-dimensional generalisation because the introduction of $\Cg$ in $\psuR$ in non-relativistic representations of massless modes should involve the same steps.  One may then endeavour to construct infinite-dimensional R-matrices that reproduce the R-matrices of \cite{Hoare:2014kma,Hoare:2014kmaa} in an IR limit yet to be determined. Another point to address is the braiding of the coproduct of \cite{Hoare:2014kma,Hoare:2014kmaa}. Even though the relativistic limit trivialises the coproduct \cite{Andrea2}, the boson $\Dg$ somehow restores the braiding in (\ref{coproduct}), hence the relationship between $\Dg$ and the braiding of the embedding superalgebra must be clarified.

It would be also interesting to make contact with the Yangian of $\psuR$ that \cite{Hoare:2014kma,Hoare:2014kmaa} put forward, under which the exact quantum S-matrix AdS$_2$  of \cite{Hoare:2014kmaa} is invariant. Based on the fact $R_{\pm}$ reproduce solution 3 and solution 5 of \cite{Andrea2} in the IR limit, one may hope to construct an embedding superalgebra that accounts for both $\algl$ and the Yangian of $\psuR$. The analysis of the so-called `secret symmetry' (see, for instance, \cite{1204.2366}) in the common embedding superalgebra may be in particular worth considering. The secret symmetry is a tower of generators that is absent at zeroth level, and whose commutation relations rises the level of other elements of the superalgebra. The action of  $\Cg_0$ is reminiscent from that of the secret symmetry, thus it may be a good starting point for this approach. We should make some remarks around this problem nonetheless. First, one may need to promote the boson $\Cg_0$ to the tower $\Cg_A$, with $A=1,2,3\dots$. Second, $\Cg_0$ does not commute with the `braiding-like' boson $\Dg$, whereas the secret symmetry does commute with braiding factors. Finally, some properties here we have assumed at level of $\algl$ may only hold at level of the representation (\ref{repinf}).

Another venue to explore are  the algebraic properties of $\algl$. First, it we would need to obtain all the close $\alg$. We have obtained the counterpart of $\Pg$, $\Kg$, and $\Hg$ in (\ref{compsu}), namely $\Pg_{A,B}$, $\Kg_{A,B}$, and $\Hg_{A,B}$ in (\ref{PKH}), for which it follows that $\algl$ closes in the representation (\ref{repinf}) due to the closure of $\alg$. However, we have not determined the closure of $\algl$ at level of the algebra, which may involve some consistency conditions. Other problem that may be worth tackling is the construction of the dual of $\alg$. The construction of the pairing of superalgebra, as \cite{1602.04988,1704.05093} proved for $\mathfrak{sl}(2|2)$,~\footnote{For recent progress on $\mathfrak{sl}(2|2)$, we refer to \cite{Matsumoto:2022nrk,Beisert:2022vnc}} involves the introduction of every outer automorphism of a given superalgebra $\algf$. Reference \cite{1602.04988,1704.05093} also showed that for the pairing to be non-degenerate, the universal enveloping algebra of $\algf$, and, hence, $\algf$ must undergo a quantum deformation. The application of these techniques to $\alg$ should shed some light on the connection between $\algl$ and Yangian superalgebras.

\section*{Acknowledgements}

The authors are grateful to Andrea Fontanella and Natalie A. Yager for comments on the manuscript. The work of J. M. N. G. has been supported by the Deutsche Forschungsgemeinschaft (DFG, German Research Foundation) under Germany's Excellence Strategy -- EXC 2121 ``Quantum Universe'' -- 39083330. The work of R. R. has been supported through the grant PGC2018-095382-B-I00, by the Universidad Complutense de Madrid (UCM) and Banco Santander through the grant GR3/14-A 910770, and by the Spanish MINECO grant PID2021-127726NB-I00. R. R. has been partially supported by UCM and Banco de Santander through the contract CT42/18-CT43/18, and partially supported by the UCM, Ministerio de Universidades, and the European Union - NextGenerationE through contract CT18/22. J. M. N. G. and R. R. are  grateful to the Fields, Strings, and Geometry Group of Department of Mathematics of the University of Surrey for kind hospitality during the completion of this work. R. R. is grateful to the Integrable Quantum Dynamics Research Group of the Department of Theoretical Physics of E{\"o}tv{\"o}s L{\'o}rand University for kind hospitality during the completion of this work. The work of A. T. has been supported by EPSRC-SFI under the grant EP/S020888/1 Solving Spins and Strings.

\end{document}